\documentclass[prc,preprint,groupedaddress,showpacs]{revtex4-1}
\usepackage{graphicx}
\newcommand{\beq}{\begin {equation}}
\newcommand{\eeq}{\end {equation}}
\begin {document}
\title{A study of nucleon-deuteron elastic scattering\\ in configuration space
}
\author{V.M. Suslov$^{1,2}$, M.A. Braun$^{1,2}$, I.N. Filikhin$^{1}$, B. Vlahovic$^{1}$ and I. Slaus$^{1,3}$}
\affiliation{$^{1}$North Carolina Central University, Durham, NC 27707, USA\\
$^{2}$Saint-Petersburg State University, 198504 St.Petersburg, Russia\\
$^{3}$R. Boskovic Institute, 10000 Zagreb, Croatia
}
\begin{abstract}
A new computational method for solving the nucleon-deuteron breakup scattering problem has been applied to study
the elastic neutron- and proton-deuteron scattering on the basis of the configuration-space Faddeev-Noyes-Noble-Merkuriev
equations. This method is based on the spline-decomposition in the angular variable and on a generalization of
the Numerov method for the hyperradius. The Merkuriev-Gignoux-Laverne approach has been generalized for
arbitrary nucleon-nucleon potentials and with an arbitrary number of partial waves. The nucleon-deuteron observables
at the incident nucleon energy 3 MeV have been calculated using the charge-independent AV14 nucleon-nucleon potential
including the Coulomb force for the proton-deuteron scattering. Results have been compared with those of other authors
and with experimental proton-deuteron scattering data.
\end{abstract}
\maketitle
\section{Introduction}
 There is an impressive amount of nucleon-deuteron scattering data: proton-deuteron and neutron-deuteron elastic
and breakup data: total, partial and differential cross sections and spin observables involving nucleon and
deuteron beams. The data are compared with the rigorous three-body theory: Faddeev-equations-based theory using
as input realistic high-precision nucleon-nucleon potentials, and including model three-nucleon forces \cite{Gloec}.
In some calculations Coulomb force has been included \cite{Alt1}. Nucleon-nucleon (NN) potentials used in rigorous
three-nucleon (3N) calculations describe the NN database with $\chi^2$/degree of freedom approximately equal to one.
These are AV18 \cite{Wir1}, CD-Bonn \cite{Mach1} and several Nijmegen potentials \cite{Stoc1} and to a lesser degree
AV14 \cite{Wir2}. Among three nucleon forces (3NF) are Tucson-Melbourne and its various modifications \cite{Coon},
and Urbana potentials \cite{Pudi}. Based on the chiral effective field theory (EFT) NN and 3N potential have been
developed \cite{Mach2} and they have been used in a rigorous 3N calculations \cite{Delt1}. A local version of the
effective field theory at next-to-next to leading order labeled ${\rm N}^2{\rm LO}$ is given in ref. \cite{Navr}.

     In spite of this enormous progress in the three-nucleon studies, there are several important cases where the
rigorous three-nucleon calculations have failed to explain the data \cite{Ivo} and these discrepancies are established
with very high precision. Among the most important discrepancies are the $\rm A_y$ puzzle in nucleon-deuteron (Nd)
elastic scattering \cite{Torn}, the star configuration in the Nd breakup reaction \cite{Wiel}, quasi-free
scattering (QFS) cross section \cite{Gloec} and the nd backward angle scattering at energies between 50 and
100 MeV \cite{How}. Some three-nucleon data show clear evidence for the 3NF, but some are in better agreement
with the calculation if the 3NF are not included. High precision realistic potentials (Nijmegen, Bonn, Paris, Urbana)
are not phase equivalent and they predict different triton binding energies, they have different short range
potentials and some differ conceptually. It is hoped that EFT will give an answer, but it is still unclear.

     There are more 3N data involving charged particles and therefore, calculations rigorously including electromagnetic
interactions are of paramount importance. The pd scattering has been studied by using hyperspherical harmonic method and
Kohn Variational Principle \cite{Kiev1} and by using the screening and renormalization procedure \cite{Alt2}. At 3 MeV
calculations have been done using high precision realistic potentials and 3NF \cite{Kiev2}, while at energies above
the threshold for the deuteron breakup only calculations using screening and renormalization have been done.
The screening method cannot be applied to energies below 1 MeV and this is a serious limitation.

     In this article we present the development of an alternative method for studying the proton-deuteron (pd) system
based on the direct numerical solution of the Faddeev-Noyes-Noble-Merkuriev (FNNM) equations in configuration space.
This approach was initiated by Merkuriev et al. (MGL) \cite{MGL} who derived general formulae for nd breakup scattering.
This method has been originally applied to study nd and pd elastic and breakup scattering but limited only to nuclear
S-waves interaction and to simple NN potentials \cite{KKM}. In the present work we generalize the MGL approach to any
high precision realistic potential for both nd and pd for elastic processes.

     This paper is organized as follows: in section 2 we describe a calculation in configuration space starting with
the general formalism in subsection 2.1, followed by Numerov method in subsection 2.2. Our novel method for solution
is given in subsection 2.3. Our results are presented in section 3. Comparisons of our results with the previous
calculations and with the data are discussed in section 4. Finally, our summary and conclusion are given in section~5.
\section {Three-nucleon Faddeev calculation in configuration space - our new computational method}
\subsection {Formalism}
The starting point for studying interactions between nucleons in three-body
systems is the solution of the Schr\"odinger equation $H\Psi = E\Psi$
for nuclear Hamiltonian such as
\beq
H = -\frac{\hbar^2}{2m}\sum_{i=1}^3\nabla^2_i + V_c + \sum_{j<k} V_{jk}\ \ \Big(+\sum_{j<k<l}
V_{jkl} \Big),
\eeq
where $V_c$ and $V_{jk}$ are the Coulomb and nuclear potentials, respectively. In this study we neglected by
three-nucleon forces $V_{jkl}$.

     Writing the total wave function as
\beq
\Psi= \Phi_1+\Phi_2+\Phi_3 = (1+P^++P^-)\Phi_1,
\eeq
the Schr\"odinger equation for three identical particles can be reduced
into a single Faddeev equation, which in Jacobi's vectors $\vec x_1,\vec y_1$
has the form
\beq
\Big[-\frac{\hbar^2}{m}\Big(\Delta_{\vec x_1}+\Delta_{\vec y_1}\Big) + V_c +
V(\vec x_1) - E\Big]\Phi(\vec x_1,\vec y_1)=-V(\vec x_1)(P^++P^-)\Phi(\vec x_1,\vec y_1),
\eeq
where the operators $P^{\pm}$ are the cyclic permutation operators for
the three particles which interchange any pair of nucleons ($ P^+:
123 \rightarrow 231, P^-: 123 \rightarrow 321$). The Coulomb potential has the
following form:
\begin{equation}
\label{vcoul}
V_c=\sum_\alpha \frac n{|x_{\alpha |}}\displaystyle \prod_{i\subset \alpha }
\frac 12(1+\tau_z^i), \ \ \ \ n=\frac{me^2}{\hbar ^2},
\end{equation}
where $e^2$=1.44 MeV $\cdot\mbox {fm}$ and $\hbar ^2/m$=41.47 MeV$\cdot\mbox
{fm}^2$. The sum runs over $\alpha =$1,2,3 for the three possible pairs and
the product of the isospin projection operators runs over the indices ${\it i}$ of the particles belonging to
the pair $\alpha$. As independent coordinates, we take the Jacobi vectors ${\overline{x}_\alpha , \overline{y}_\alpha }$.
For the pair $\alpha $=1, they are related to particle coordinates by the formulas:
\begin{equation}
\overline{x}_1=\overline{r}_2-\overline{r}_3, \ \ \ \ \ \ \overline{y}_1 =%
\frac{\overline{r}_2+\overline{r}_3}2 -\overline{r}_1,
\end{equation}
for $\alpha $=2,3 one has to make cyclic permutations of the indexes in Eq.(5). The Jacobi vectors with
different $\alpha$'s are linearly related by the orthogonal transformation
\begin{equation}
  \left(
  \begin{array}{c}
     \overline{x}_{\alpha} \\ \overline{y}_{\alpha}
  \end{array}
  \right)=
  \left(
  \begin{array}{rl}
      C_{\alpha\beta} & S_{\alpha\beta} \\
     -S_{\alpha\beta} & C_{\alpha\beta}
  \end{array}
  \right)
  \left(
  \begin{array}{c}
     \overline{x}_{\beta} \\ \overline{y}_{\beta}
  \end{array}
  \right) \ ,\ \ \ C^2_{\alpha\beta} + S^2_{\alpha\beta} = 1,
\end{equation}
where
\begin{equation}
C_{\alpha\beta}=-\sqrt{\frac{m_{\alpha}m_{\beta}}
{(M-m_{\alpha})(M-m_{\beta})}}, \ \
S_{\alpha\beta} = (-)^{\beta - \alpha}{\rm sgn}(\beta - \alpha)
\sqrt{1-C^{2}_{\alpha\beta}}, \ \ M=\sum_{\alpha=1}^3m_{\alpha}.
\end{equation}

To perform numerical calculations for arbitrary nuclear potential, we use MGL approach \cite{MGL}. For $pd$ scattering
the FNNM equations for partials components can be written in the following form (here we omit the index 1):
\begin{equation}
\label{fexy}
\begin{array}{c}
\Big[E+\frac{\hbar^2}{m}(\partial_x^2+\partial_y^2)-v_{\alpha}^{\lambda l}(x,y)
\Big]\Phi^{\lambda_0,s_0,M_0}_{\alpha}(x,y) =
\sum_{\beta}\Big[v_{1,\alpha\beta}+\sum_{\tau}(v^{+}_{\tau}
c^{M_0+}_{\tau,\alpha\beta} + v^{-}_{\tau}c^{M_0-}_{\tau,\alpha\beta})
\Big]
\\
\\
\times\Phi^{\lambda_0,s_0,M_0}_{\beta}(x,y)
+\sum_{\beta}v_{\alpha\beta}(x)\Big[\Phi^{\lambda_0,s_0,M_0}_{\beta}(x,y)+
\int_{-1}^1du\sum_{\gamma}g_{\beta\gamma}(y/x,u)
\Phi^{\lambda_0,s_0,M_0}_{\gamma}(x',y')\Big].
\end{array}
\end{equation}
Here Greek subindexes denote state quantum numbers: $\alpha=\{l,\sigma,j,s,\lambda,t,T\}$,
where $l$, $\sigma$, $j$ and $t$ are the orbital, spin, total angular momenta and isospin of a pair of
nucleons, $\lambda$ is the orbital momentum of the third nucleon relative to the c.m.s. of a pair nucleons,
and $s$ is the total "spin" (${\bf s =1/2 + j}$). ${\bf M = \vec \lambda + \vec s}$ is the total three-particle
angular momentum, and the value of total isospin is $T$. In Eqs.(\ref{fexy}) $v_1$ and coefficients $c^{M_0\pm}$
depending on quantum state numbers of channel combined with $v^{\pm}$ are matrix elements of the Coulomb potential
projected onto the MGL basis. For given $\alpha$ and $\beta$ summation over $\tau$ in Eqs.(\ref{fexy}) is finite.
If $\alpha=\{l \sigma j s \lambda t T\}$ and $\beta=\{l'\sigma' j' s' \lambda' t' T'\}$ then values of $\tau$
are restricted by the following inequality: $$\max(|l-l'|,\lambda-\lambda'|)\leq\tau\leq \min(l+l',\lambda+\lambda')$$.
This means that for a chosen set of basic states, Eqs.(\ref{fexy}) take into account the Coulomb interaction exactly
(although the latter has been expanded in partial waves).

     The geometrical function $g_{\beta\gamma}(x,y,u)$ is the representative of the permutation operator $P^++P^-$
in MGL basis \cite{MGL}:
\[\
g_{\alpha'\alpha}(y/x,u)=g_{\alpha'\alpha}(\theta,u)=
g_{\alpha'\alpha}(\theta,\theta')\]\[=
(-1)^{\lambda+\lambda'+J+J'}
[(2J+1)(2J'+1)(2s+1)(2s'+1)]^{1/2}
\sum_{LS}(2S+1)(2L+1)\]\[
    \left \{ \begin{array}{ccc}
    l     & \sigma & J \\
    1/2 & s & S
    \end{array} \right \}
 \left \{ \begin{array}{ccc}
    l'     & \sigma' & J' \\
    1/2 & s' & S
    \end{array} \right \}
 \left \{ \begin{array}{ccc}
    \lambda     & l & L \\
    S & M & s
    \end{array} \right \}
 \left \{ \begin{array}{ccc}
    \lambda'     & l' & L \\
    S & M & s'
    \end{array} \right \}\]\beq
<\chi^S_{1/2\sigma'}\eta^T_{1/2,t'}|P^+|\chi^S_{1/2\sigma}\eta^T_{1/2,t}>
h^L_{\lambda'l'\lambda l}(y/x,u).
\eeq

Function $h$ is the representative of the permutation operator
$P^++P^-$ in the $\lambda+ l=L$ basis:
\[ h^L_{\lambda'l'\lambda l}(y/x,u)=h^L_{\lambda'l'\lambda
l}(\theta,u)= h^L_{\lambda'l'\lambda l}(\theta,\theta')\]\[=
\frac{xy}{x'y'}(-1)^{l+L}\frac{(2\lambda+1)(2l+1)}{2^{\lambda+l}}
[(2\lambda)!(2l)!(2\lambda'+1)(2l'+1)]^{1/2}\]\[
\sum_{k=0}(-1)^k(2k+1)P_k(u)\sum_{\lambda_1+\lambda_2=\lambda,\,
l_1+l_2=l}\frac{y^{\lambda_1+l_1}x^{\lambda_2+l_2}}{{y'}^{\lambda}{x'}^l}
(-1)^{l_2}\]\[\frac{(\sqrt{3})^{\lambda_2+l_1}}
{[(2\lambda_1)!(2\lambda_2)!(2l_1)!(2l_2)!]^{1/2}}
\sum_{\lambda''l''}(2\lambda''+1)(2l''+1) \]\[\left (
\begin{array}{ccc}
    \lambda_1     & l_1 & \lambda''\\
    0 & 0 & 0
    \end{array} \right )
\left ( \begin{array}{ccc}
    \lambda_2     & l_2 & l''\\
    0 & 0 & 0
    \end{array} \right )
\left ( \begin{array}{ccc}
    k     & \lambda'' & \lambda'\\
    0 & 0 & 0
    \end{array} \right )
\left ( \begin{array}{ccc}
    k     & l'' & l'\\
    0 & 0 & 0
    \end{array} \right )\]\beq
\left \{ \begin{array}{ccc}
    l'     & \lambda' & L\\
    \lambda'' & l'' & k
    \end{array} \right \}
\left \{ \begin{array}{ccc}
    \lambda_1     & \lambda_2 & \lambda\\
    l_1 & l_2 & l\\
    \lambda''&l''&L
    \end{array} \right \}.
\eeq
The index $k$ runs from zero to $(\lambda'+l'+\lambda+l)/2$.
The $(...)$ are the 3$j$ symbols:
\[
\left ( \begin{array}{ccc}
    j_1     & j_2 & j_3\\
    m_1 & m_2 & m_3
    \end{array} \right )= (-1)^{j_3+m_3+2j_1}\frac{1}{\sqrt{2j_3+1}}
C^{j_3m_3}_{j_1-m_1j_2-m_2}.
\]
The centrifugal potential is
\beq
v^{\lambda l}_{\alpha}=\frac{\hbar^2}{m} \Big[\frac{l(l+1)}{x^2}+
\frac{\lambda(\lambda+1)}{y^2} \Big],
\eeq
and nucleon-nucleon potentials are $v_{\alpha\alpha'}(x)=<\alpha|v({\bf
x})|\alpha'>=\delta_{\lambda\lambda'}
\delta_{ss'}\delta_{\sigma\sigma'}\delta_{JJ'}v^{\sigma J}_{ll'},$
where $v^{\sigma J}_{ll'}$ are the potential representatives in the two-body basis ${\cal Y}^{JJ_z}_{l\sigma}
({\bf\hat{x}})$ (most often abbreviated as $^{2\sigma+1}l_J)$.

    The set of partial differential equation Eqs.(\ref{fexy}) must be solved for functions satisfying
the regularity conditions
\begin{equation}
\label{rcond}
\Phi^{\lambda_0 s_0 M_0}_{\alpha}(0,\theta)=\Phi^{\lambda_0 s_0 M_0}_{\alpha}(\rho,0)
=\Phi^{\lambda_0 s_0 M_0}_{\alpha}(\rho,\pi/2)=0
\end{equation}

    The asymptotic conditions for pd elastic scattering has the following form \cite{MerkAs}:
\begin{equation}
\label{asympt}
\begin{array}{c}
\Phi^{\lambda_0s_0M_0}_{1,\bar{\alpha}}(x,y)\sim
\Big\{\delta_{\lambda\lambda_0}\delta_{ss_0}\delta_{\sigma 1}
\delta_{j1}e^{i\Delta^c_\lambda}F_\lambda^c(qy) +
e^{-i\Delta_\lambda^c}\Big(G^c_\lambda(qy)+iF^c_\lambda(qy)\Big)
a^{M_0}_{\lambda s\lambda_0 s_0}\Big\} \psi_l(x),
\\
\\
 \ \ x\ {\rm finite},\ \ y\to\infty ,
\end{array}
\end{equation}
where $\Delta_\lambda^c=$arg$\Gamma(\lambda+1+i\nu)$ is the Coulomb phase and $\nu$ is equal $n/(\sqrt 3q)$,
$\psi_l$ is $l-th$ component of deuteron wave function ($l=$0,2), and $F^c$ and $G^c$ are the regular and
irregular Coulomb functions, respectively.

     The S-matrix is defined as follows
\begin{equation}
S^{M_0}_{\lambda s\lambda_0 s_0}=\delta_{\lambda\lambda_0}
\delta_{ss_0}\delta_{\sigma 1}\delta_{j1}e^{i2\Delta^c_\lambda}
+2ia^{M_0}_{\lambda s\lambda_0 s_0}.
\end{equation}
At energies below threshold the $S$-matrix is unitary and may be presented as
$$S=e^{2i\Delta},$$ where $\Delta$ is the Hermitean matrix of scattering phases.
From (\ref{asympt}) we then find that the matrix of partial elastic amplitudes
$a$ has the structure
\begin{equation}
a=\frac{\Big(e^{2i\Delta}-e^{2i\Delta^c} \Big)}{2i}
=\frac{e^{2i\Delta^c}\Big(e^{2i\delta}-1\Big)}{2i},
\end{equation}
where $\Delta^c$ is a diagonal matrix of Coulomb phases and $\Delta$ is the Hermitean matrix of scattering phases.
The phases $\delta =\Delta-\Delta^c$ are the contribution to the scattering phase due to the nuclear interaction.

     To simplify the numerical solution the FNNM equations, we write down Eqs.(\ref{fexy}) in the polar coordinate
system ($\rho^{2} = x^{2}+y^{2}$ and $\tan\theta = y/x$):
\begin{equation}
\label{fe}
\begin{array}{c}
\displaystyle
\Big[E+\frac{\hbar^2}{m}(\frac{\partial^2}{\partial\rho^2}
+\frac1{\rho^2}\frac{\partial^2}{\partial\theta^2}+\frac1{4\rho^2})
-v_{\alpha}^{\lambda l}(\rho,\theta)
\Big]U^{\lambda_0 s_0 M_0}_{\alpha}(\rho,\theta) = \frac{n}{\rho}\sum_{\beta}
Q_{\alpha\beta}U^{\lambda_0 s_0 M_0}_{\beta}(\rho,\theta)
\\
\\
\displaystyle
+\sum_{\beta}v_{\alpha\beta}(\rho,\theta)\Big[U^{\lambda_0 s_0 M_0}_{\beta}(\rho,\theta)+
\int_{-1}^1du\sum_{\gamma}g_{\beta\gamma}(\theta,u,\theta'(\theta,u))
U^{\lambda_0 s_0 M_0}_{\gamma}(\rho,\theta')\Big],
\end{array}
\end{equation}
where
\begin{equation}
\cos ^2\theta ^{^{\prime }}(u,{\theta })=\frac 14\cos ^2\theta - \frac{\sqrt{%
3}}2\cos \theta \sin \theta \cdot u+\frac 34\sin ^2\theta,
\end{equation}
and the first derivative in the radius is eliminated by the substitution $U = \rho^{-1/2}\Phi$. In Eq. (\ref{fe}) $
 Q_{\alpha\beta}$ is the overall matrix sum of the Coulomb potential.

     In the case of neutron-deuteron elastic scattering one has to set the "charge" $n$ equal to zero. This leads
to equality to zero of the Coulomb phases $\Delta^c_{\lambda}$, and the Coulomb functions $F^c_{\lambda}$ and
$G^c_{\lambda}$ are reduced to the regularized spherical Bessel functions $\hat j_{\lambda}$ and $-\hat y_{\lambda}$,
respectively.

\subsection {Numerov method}
   Modification of the Numerov method for the set of the differential
equations (\ref{fe}) does not present any difficulties in principle.
As is well known, the Numerov method is an efficient algorithm for solving
second-order differential equations. The important feature of the equations
for the application of Numerov's method is that the first derivative has to
be absent. The aim of this method is to improve the accuracy
of the finite-difference approximation for the second derivative.
Starting from the Taylor expansion
truncated after the sixth derivative
for two points adjacent to $x_n$, that is for $x_{n-1}$ and
$x_{n+1}$ one sums these two expansions to give a new computational
formula that includes the fourth derivative. This derivative can be
found by straightforward differentiation of the second derivative
from the initial second-order differential equation
(see the details in \cite{Sus}). For brevity, we omit the
corresponding derivation and present only the final formula of
Numerov's method for the FNNM equations (omitting the upper indices $\lambda_0s_0M_0 $):
\begin{equation}
\label{reseq}
\begin{array} {l}
\displaystyle
-\Big[E + \frac{12}{(\Delta\rho)^2}+(1+\frac{2\Delta\rho}{\rho_j})\frac{T_{\alpha}(\theta)}{\rho^2_j}\Big]
U_{\alpha}(\rho_{j-1},\theta)
+n\sum_{\beta}\frac{Q_{\alpha\beta}(\theta)}{\rho_j}(1+\frac{\Delta\rho}{\rho_j})U_{\beta}(\rho_{j-1},\theta)
\\
\displaystyle
+\sum_{\beta}(v_{\alpha\beta}(\rho_j,\theta)-\Delta\rho v'_{\alpha\beta}(\rho_j,\theta))(U_{\beta}(\rho_{j-1},\theta)
+\sum_{\gamma}\int_{\theta^-}^{\theta^+}d\theta^{\prime}g_{\beta\gamma}(\theta,\theta')
U{\gamma}(\rho_{j-1},\theta^{\prime}))
\\
\displaystyle
-2\Big[5E - \frac{12}{(\Delta\rho)^2}+(5+\frac{3\Delta\rho}{\rho_j})\frac{T_{\alpha}(\theta)}{\rho^2_j}\Big]
U{\alpha}(\rho_j,\theta)
+2n\sum_{\beta}\frac{Q_{\alpha\beta}(\theta)}{\rho_j}(5+\frac{(\Delta\rho)^2}{\rho^2_j})U_{\beta}(\rho_j,\theta)
\\
\displaystyle
+\sum_{\beta}(10v_{\alpha\beta}(\rho_j,\theta)+(\Delta\rho)^2v^{\prime\prime}_{\alpha\beta}(\rho_j,\theta))
(U_{\beta}(\rho_j,\theta)+\sum_{\gamma}\int_{\theta^-}^{\theta^+}d\theta^{\prime}g_{\beta\gamma}(\theta,\theta')
U_{\gamma}(\rho_j,\theta^{\prime}))
\\
\displaystyle
-\Big[E+ \frac{12}{(\Delta\rho)^2}+(1-\frac{2\Delta\rho}{\rho_j})\frac{T_{\alpha}(\theta)}{\rho^2_j}\Big]
U_{\alpha}(\rho_{j+1},\theta)
+n\sum_{\beta}\frac{Q_{\alpha\beta}(\theta)}{\rho_j}(1-\frac{\Delta\rho}{\rho_j})U_{\beta}(\rho_{j+1},\theta)
\\
\displaystyle
+\sum_{\beta}(v_{\alpha\beta}(\rho_j,\theta)+\Delta\rho v'_{\alpha\beta}(\rho_j,\theta))(U_{\beta}(\rho_{j+1},\theta)
+\sum_{\gamma}\int_{\theta^-}^{\theta^+}d\theta^{\prime}g_{\beta\gamma}(\theta,\theta')
U_{\gamma}(\rho_{j+1},\theta^{\prime})),
\end{array}
\end{equation}
where
$$T_{\alpha}(\theta)=\frac{\partial^2}{\partial\theta^2}-\frac{l(l+1)}{\cos^2\theta}
-\frac{\lambda(\lambda+1)}{\sin^2\theta}+\frac14.$$
In Eq. (\ref{reseq}) $\rho_j$ is the $j-th$ current point for hyperradius $\rho \in (0,R_{max})$ in the radial
grid ($j=1,2,\dots , N_{\rho}$), $\Delta\rho_j$ is the radial step-interval.

     To ensure the accuracy of order $(\Delta \theta)^4$ for the approximation in the angular variable, Hermitian
splines of the fifth degree have been used (see Ref. \cite{KviH}). These splines are local and each spline
$S_{\sigma i}(x)$ is defined for $x$ belonging to two adjacent subintervals $[x_{i-1},x_i]$ and $[x_i,x_{i+1}]$.
Their analytical form is fixed by the following smoothness conditions:
\begin{equation}
S_{\sigma i}(x_{i-1})=0,\ \ S_{\sigma i}(x_{i+1})=0, \ \ \sigma=0,1,2,
\end{equation}
and
\begin{equation}
\begin{array}{l}
S_{0i}(x_i)=1, \ \ \ \ S^{\prime}_{0i}(x_i)=0, \ \ \ \ S^{\prime\prime}_{0i}(x_i)=0,\\
S_{1i}(x_i)=0, \ \ \ \ S^{\prime}_{1i}(x_i)=1, \ \ \ \ S^{\prime\prime}_{1i}(x_i)=0,\\
S_{2i}(x_i)=0, \ \ \ \ S^{\prime}_{2i}(x_i)=0, \ \ \ \ S^{\prime\prime}_{2i}(x_i)=1.\\
\end{array}
\end{equation}
Expansion of the Faddeev component into basis of the Hermitian splines has the following form:
\begin{equation}
\label{sex}
U_{\alpha}(\rho,\theta)=\sum_{\sigma=0}^{2}\sum_{j=0}^{N_{\theta}+1}S_{\sigma j}(\theta)
C^{\sigma}_{\alpha j}(\rho),
\end{equation}
where $N_{\theta}+1$ is the number of internal subintervals for the angular variable $\theta \in [0,\pi/2] $.

     To reduce the resulting equation (\ref{reseq}) to an algebraic problem, one should explicitly calculate
the derivatives of $NN$ potentials $v_{\alpha\beta}(\rho,\theta)$ with respect to $\rho$ and the second derivates
of splines $S_{\sigma j}(\theta)$ with respect to $\theta$. It is convenient to express the second derivative of
component $U_{\alpha}$ with respect to $\theta$ through $U_{\alpha}$ itself using Eq.(\ref{sex}). Upon substituting
the spline expansion (\ref{sex}) and expression for its second derivative into Eqs.(\ref{reseq}), we use a
collocation procedure with three Gaussian quadrature points per subinterval. As the number of internal breakpoints
for angular variable $\theta$ is equal to $N_{\theta}$, the basis of quintic splines consists of $3N_{\theta}+6$
functions. Three of them should be excluded using the last two regularity conditions from (\ref{rcond}) and
continuity of the first derivative in $\theta$ of the Faddeev component at either $\theta = 0$ or $\theta = \pi/2$,
as the collocation procedure yields $3N_{\theta}+3$ equations. Finally Eqs.(\ref{reseq}) for the Faddeev components
are to be written as the following matrix equation:
\begin{equation}
\label{algpr}
\begin{array}{l}
 \ \ \ \ \ \ \ \ \ \ \ \ \ \ \ \ \ \ A_{1}U_{1}+G_{1}U_{2}\ \  =0,\\
B_{j}U_{j-1}\ \ \ \ +A_{j}U_{j}+G_{j}U_{j+1}=0,\ \ \ \ \ \ \ \ \ \ \ \ \ \ j=2,...N_{\rho}-1,\\
B_{N_{\rho}}U_{N_{\rho}-1}+A_{N_{\rho}}U_{N_{\rho}}\ \ \ \ \ \ \ \ \ \ =-G_{N_{\rho}}U_{N_{\rho}+1}.
\end{array}
\end{equation}
In this equation vector $U_{k} = U(\rho_{k})$ has dimension $N_{in}$ and matrices  $B,A,G$ have dimension
$N_{in} \times N_{in}$ where $N_{in}$ = $N_{\alpha} \times N_c$, and $N_{\alpha}$ is the number of partial
waves and $N_c=3N_{\theta}+3$ is the number of collocation points in the angular variable $\theta$.
\subsection {The novel method of solution}
To derive equations for calculation of elastic Nd amplitudes, the method of
partial inversion \cite{Sus} has been applied. We write down Eq.(\ref{algpr})
in a matrix form:
\begin{equation}
\label{mateq}
(D*U)_i=-\delta_{iN_{\rho}}G_{N_{\rho}}U_{N_{\rho}+1}.
\end{equation}
Here matrix D is of dimension $N_{\rho}N_{in}\times N_{\rho}N_{in}$, and $N_{\rho}$ is the number of breakpoints
in the hyperradius $\rho$. The form of this equation results from keeping the incoming wave in the asymptotic
condition (\ref{asympt}). As a consequence, the right hand part of Eq.(\ref{mateq}) has a single nonzero term
marked with index $N_{\rho}+1$. Sparse (tri-block-diagonal) structure of matrix D optimizes considerably the
inversion problem.

      Hyperradius $\rho_{N_{\rho}+1}=R_{max}$, where $R_{max}$ is the cutoff radius at which the asymptotic
condition Eq.(\ref{asympt}) are implemented. By formal inversion of the matrix D in Eq. (\ref{mateq}), the solution
of the problem may be written in the following form:
\begin{equation}
\label{solv}
  U_{j}=-D^{-1}_{jN_{\rho}}G_{N_{\rho}}U_{N_{\rho}+1},  \ \ \ \ j=1,2....N_{\rho}.
\end{equation}
In Eqs.(\ref{solv}) one should consider the last component of vector $U$:
\begin {equation}
U_{N_{\rho}} = -D^{-1}_{N_{\rho}N_{\rho}}G_{N_{\rho}}U_{N_{\rho}+1}.
\end {equation}
Provided $R_{max}$ is large enough, the vector $U_{N_{\rho}}$ on the left side
of Eq. (25) may be replaced by the corresponding vector obtained by evaluating
Eq. (\ref{asympt}) at the radius $\rho = \rho_{N_{\rho} }$. As a result in the case $M \geq 3/2$ we obtain
three linear equations for the unknown amplitudes
${a}_{\lambda s \lambda_0s_0}^{M_0}$:
\begin{equation}
\label{ipp}
\sum_{i=1}^3 a_{ij}^{M_0}\cdot {\bf v}^i  ={\bf F}^j, \ \ \ \ j=1,2,3.
\end{equation}
For $M_0$= 1/2 the indices run over ${i,j}$=1,2. In these equations indices $i,j$ number the asymptotic values
of pairs ($\lambda s$), and vectors $\bf v,{\bf F}$ are known quantities.
For the sake of brevity, we do not display here the explicit form of them. As $R_{max} \rightarrow \infty$
the set of equations (\ref{ipp}) has a set of constants $a_{ij}^{M_0}$ as a solution. At finite $R_{max}$ its
solution is a vector $a$ with generally different components corresponding to different angles.

     For each value of $j$ linear equation (\ref{ipp}) is over determined, since the number of equations is
$N_{in}$ and the number of unknowns is $3$. Therefore it is natural to use the least-squares method (LSM) as
was proposed in \cite{Sus}. According to LSM one has to minimize the following functional
\begin{equation}
\|\sum_{i=1}^3 a_{ij}^{M_0}\cdot {\bf v}^i-{\bf F}^j\|^2=\min.
\end{equation}

Differentiating this expression with respect to Re\,$a_{ij}^{M_0} $ and Im\,$a_{ij}^{M_0}$ we obtain
three(two) sets of liner complex equations of dimension $3 \times 3$ ($2 \times 2$ for $M_0$=1/2), respectively.
\begin {equation}
  \sum_{i=1}^3 a_{ij}^{M_0}\cdot({\bf v}^{\ast}_k,{\bf v}^i)
  = ({\bf v}^{\ast}_k,{\bf F}^j),\ \ j=1,2,3, \ \ k=1,2,3,
\end {equation}
where $(\xi^{\ast},f)=\sum_{i}\xi^{\ast}_i\cdot f_i$ is an ordinary scalar product. Now calculation of
amplitudes $a_{ij}^{M_0}$ is trivial task.
\subsection {Observables}
     To calculate observables for elastic scattering of nucleon
from deuteron in the direction ${\bf\hat{q}'}$ (initial direction ${\bf\hat{q}}$
is along the z-axis), one has to derive the equation for the elastic amplitude as a function of
scattering angle. Omitting this derivation, we represent the final expression for this amplitude in MGL basis:
\begin{equation}
\label{hata}
\begin{array}{c}
\displaystyle
\hat{a}_{\sigma'_z,J'_z,\sigma_z,J_z}({\bf\hat{q}'}) =
\sum_{M}\sum_{\lambda' s'}\sum_{\lambda s}i^{\lambda-\lambda'}\sqrt{\frac{2\lambda+1}{4\pi}}
\\
\displaystyle
C^{MM_z}_{\lambda' M_z-\sigma'_z-J_z',s' \sigma_z'+J_z'}
C^{MM_z}_{\lambda 0, s \sigma_z+J_z}C^{s' \sigma_z'+J_z'}_{1/2 \sigma_z', 1 J_z'}
C^{s \sigma_z+J_z}_{1/2 \sigma, 1 J_z}Y_{\lambda' M_z-\sigma_z'-J_z'}(
{\bf\hat{q'}})a^M_{\lambda' s' \lambda s},
\end{array}
\end{equation}
with $M_z=\sigma_z+J_z$.

     In Eq. (\ref{hata}) $\sigma' \sigma_z' (\sigma, \sigma_z)$ and $J' J_z'( JJ_z)$ are spin
and its projection for incoming (scattered) nucleon, and the deuteron in the rest (scattered deuteron), respectively.
Thus, the nuclear part of the elastic amplitude is a $(2\times 2)\otimes(3\times 3)$ matrix in the spin states
of nucleon and deuteron, depending on the spherical angles $\theta$ and $\phi$.

     The situation is a little bit more complicated with the elastic scattering of proton from deuteron,
since apart from the nuclear part the elastic amplitude also contains the pure Coulomb part. Thus in the matrix
notation the resulting amplitude is to be sum of two amplitudes:
\begin{equation}
\hat{a}^{tot}=\hat{a}+\hat{a}^c,
\end{equation}
where $\hat{a}$ is the nuclear part of the same form as for the nd case and $\hat{a}^c$ is the Coulomb part
which is a unit matrix in spin states (this term does not change spins and depends only on $\theta$):
\begin{equation}
\hat{a}^c_{\sigma'_z,J'_z,\sigma_z,J_z}({\bf\hat{q}'})=
a^c(\theta)\delta_{\sigma'_z\sigma_z}\delta_{J'zJ_z}.
\end{equation}
The amplitude $a^c$ is as follows
\begin{equation}
a^c(\theta)=-\frac{n}{8\pi q\sqrt3\sin^2 (\theta/2)}
e^{-i\nu\ln\sin^2(\theta/2)+2i\eta^c}.
\label{oura}
\end{equation}
The parameter $\nu$ is defined by the ratio $\nu=\frac{2n}{3q}$,
$q$ is the wave vector of proton, the parameter $n$ is given in Eq. (\ref{vcoul}), and $\eta^c=$ arg $\Gamma(1+i\nu)$.

     The spin observable formulas can be taken from the review of W. Gl\"ockle et
al. \cite{Gloec}. They are expressed via spin $2\times 2$ matrices $\sigma_i$ for the nucleon and
$3\times 3$ matrices ${\cal P}_i$ and ${\cal P}_{ik}$ for the deuteron. The latter are related to the
deuteron spin matrices $S_i$:
 \beq
S_x=\frac{1}{\sqrt{2}}\left ( \begin{array}{ccc}
    0     & 1 & 0\\
    1 & 0 & 1\\
    0& 1& 0
    \end{array} \right),\
S_y=\frac{1}{\sqrt{2}}\left ( \begin{array}{ccc}
    0     & -i & 0\\
    i & 0 & -i\\
    0& i& 0
    \end{array} \right),\
S_z=\left ( \begin{array}{ccc}
    1     & 0 & 0\\
    0 & 0 & 0\\
    0& 0& -1
    \end{array} \right).\
\eeq
One has ${\cal P}_i=S_i$, ${\cal P}_{ik}=3/2(S_iS_k+ S_kS_i)$, ${\cal P}_{zz}=3S_zS_z-2I$, and
${\cal P}_{xx}-{\cal P}_{yy}=3(S_xS_x- S_yS_y)$.

Nucleon analyzing powers $A_k$ are
 \beq
 A_k=\frac{{\rm
Tr}\,(\hat{a}\sigma_k\hat{a}^{\dagger})} {{\rm Tr}\,(\hat{a}\hat{a}^{\dagger})}.
\eeq
If the scattering plane is the $xy$ plane and the $y$ axis points to the direction $\bf{q}\times\bf{q'}$ then
due to parity conservation $A_x=A_z=0$ and the only non-zero component is $A_y$.

The deuteron vector and tensor analyzing powers are defined as
\beq
A_k=\frac{{\rm Tr}\,(\hat{a}{\cal P}_k\hat{a}^{\dagger})} {{\rm
Tr}\,(\hat{a}\hat{a}^{\dagger})},\ A_{jk}=\frac{{\rm
Tr}\,(\hat{a}{\cal P}_{jk}\hat{a}^{\dagger})} {{\rm
Tr}\,(\hat{a}\hat{a}^{\dagger})}.
 \eeq
Parity conservation puts $A_x,A_z,A_{xy}$ and $A_{yz}$ to zero. So the non-vanishing and independent analyzing
powers are defined by
 \beq
iT_{11}=\frac{\sqrt{3}}{2}A_y,\ \ T_{20}=\frac{1}{\sqrt{2}}A_{zz},\
\ T_{21}=-\frac{1}{\sqrt{3}}A_{xz},\ \
T_{22}=\frac{1}{2\sqrt{3}}(A_{xx}-A_{yy}).
\eeq

Also spin transfer coefficients are given in the review. They have
the same structure as the quantities above, with slightly different
matrices to be inserted between $\hat{a}$ and $\hat{a}^{\dagger}$.

\section {Results}
\begin{figure}[h]
\vfill
\begin{minipage}[h]{100mm}
\includegraphics[width=100.0mm]{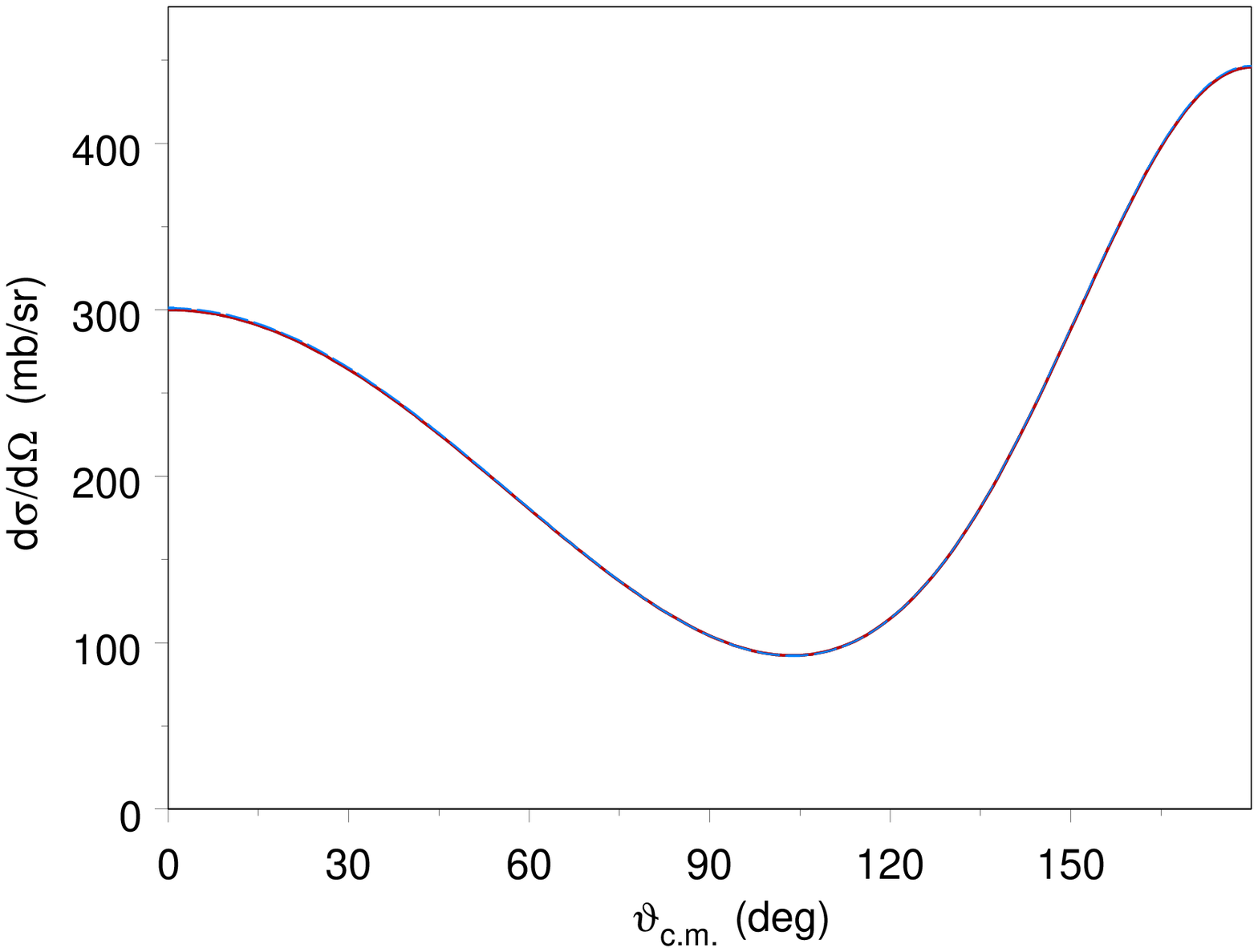}
\end{minipage}
\begin{minipage}[h]{100mm}
\includegraphics[width=100mm]{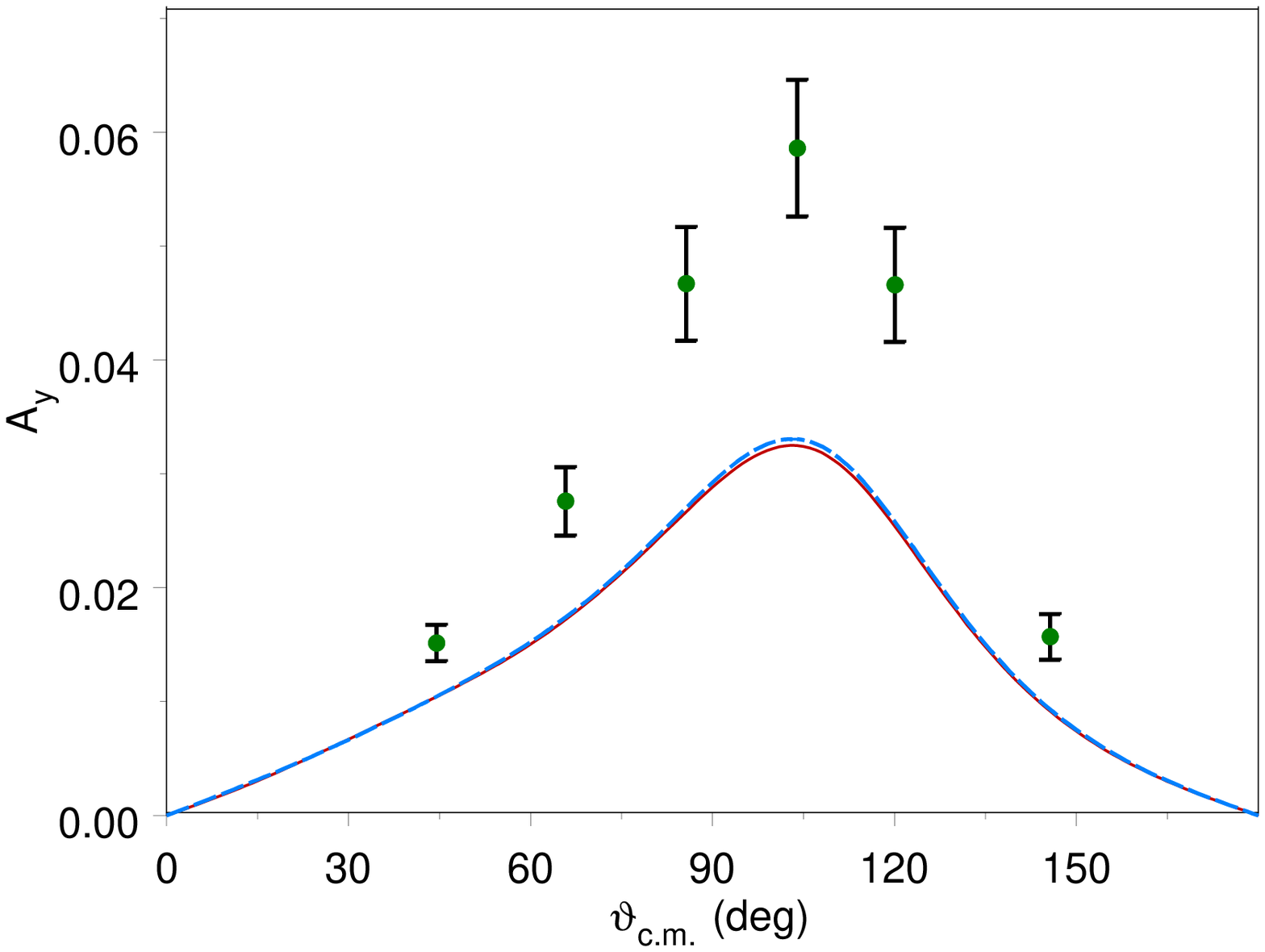}
\end{minipage}
\caption {Differential cross section and neutron analyzing power $\rm A_y$ for nd elastic scattering at 3~MeV lab energy
as function of the c.m. scattering angle. The solid lines correspond to our results obtained with AV14 NN potential.
The dashed lines correspond to results of Kievsky at al. obtained with AV14 NN potential \cite{KievG}. The experimental
data are from Ref. \cite{Anin}.}
\label{fig:1}
\end{figure}

\begin{figure}[h]
\vfill
\begin{minipage}[h]{100mm}
\includegraphics[width=100mm]{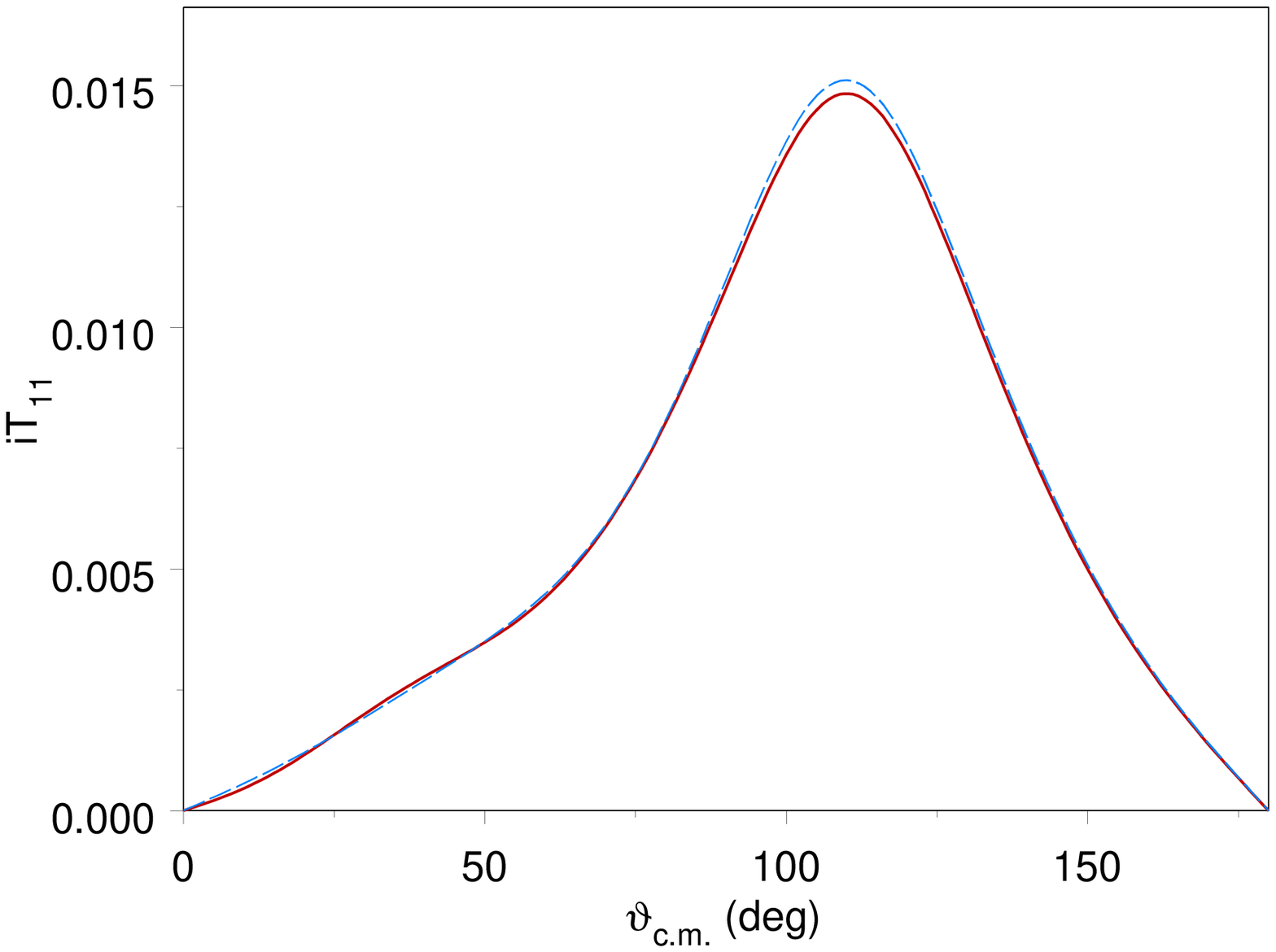}
\end{minipage}
\begin{minipage}[h]{100mm}
\includegraphics[width=100mm]{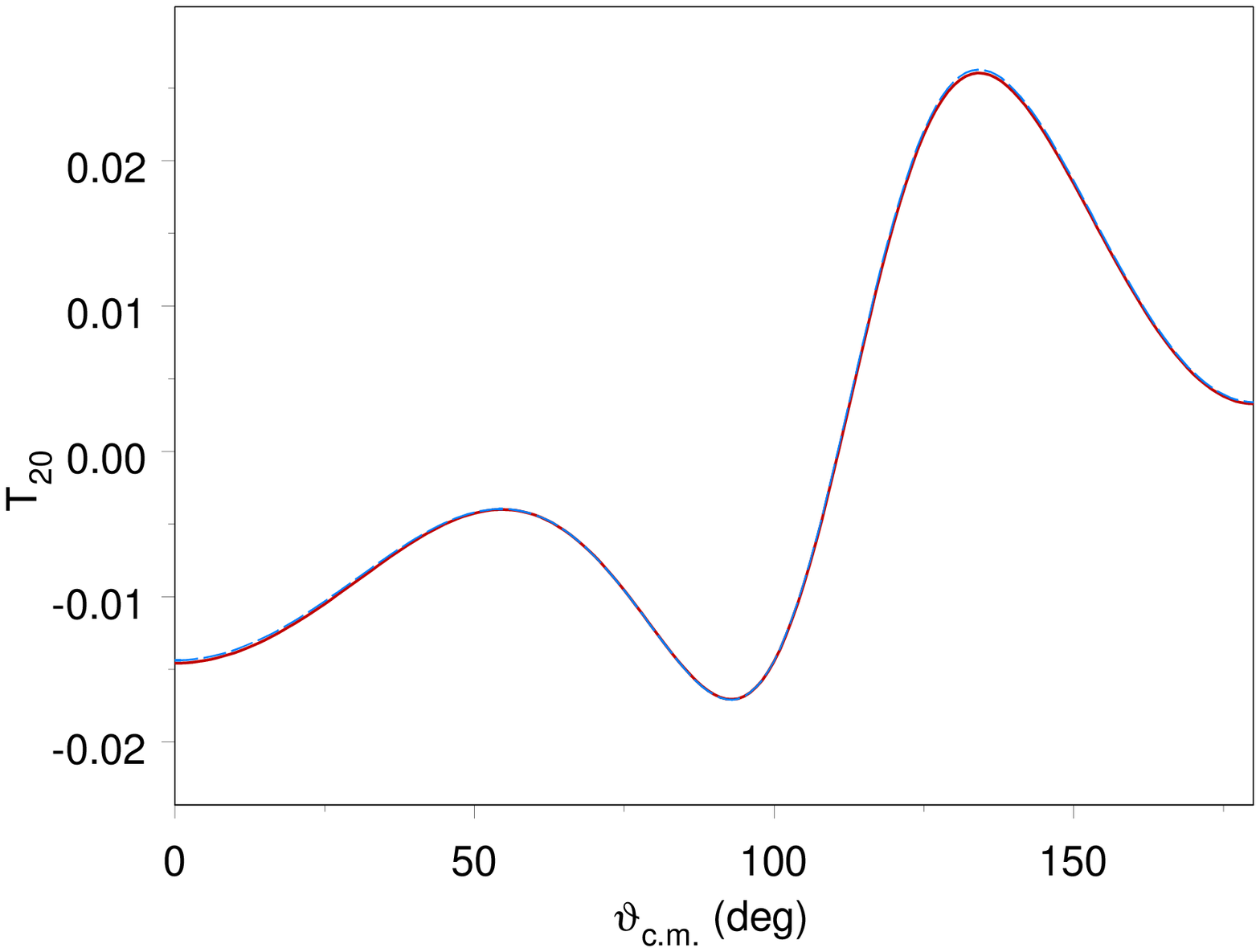}
\end{minipage}
\caption{Deuteron vector $i\rm T_{11}$ and tensor $\rm T_{20}$ analyzing power for nd elastic scattering at 3~MeV lab
energy as function of the c.m. scattering angle. The notations are the same as in Fig. \ref{fig:1}.}
\label{fig:2}
\end{figure}

\begin{figure}[h]
\vfill
\begin{minipage}[h]{100mm}
\includegraphics[width=100mm]{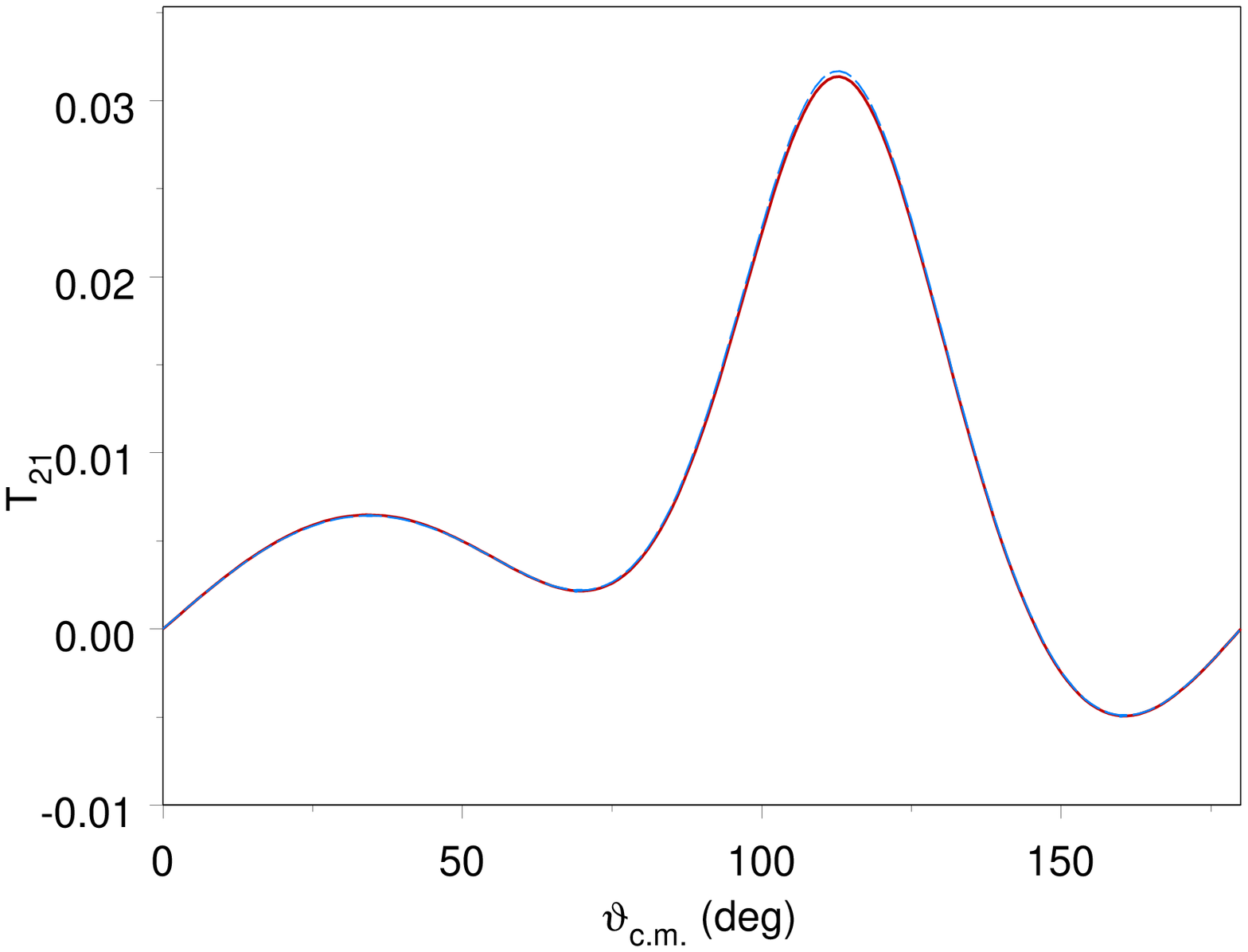}
\end{minipage}
\begin{minipage}[h]{100mm}
\includegraphics[width=100mm]{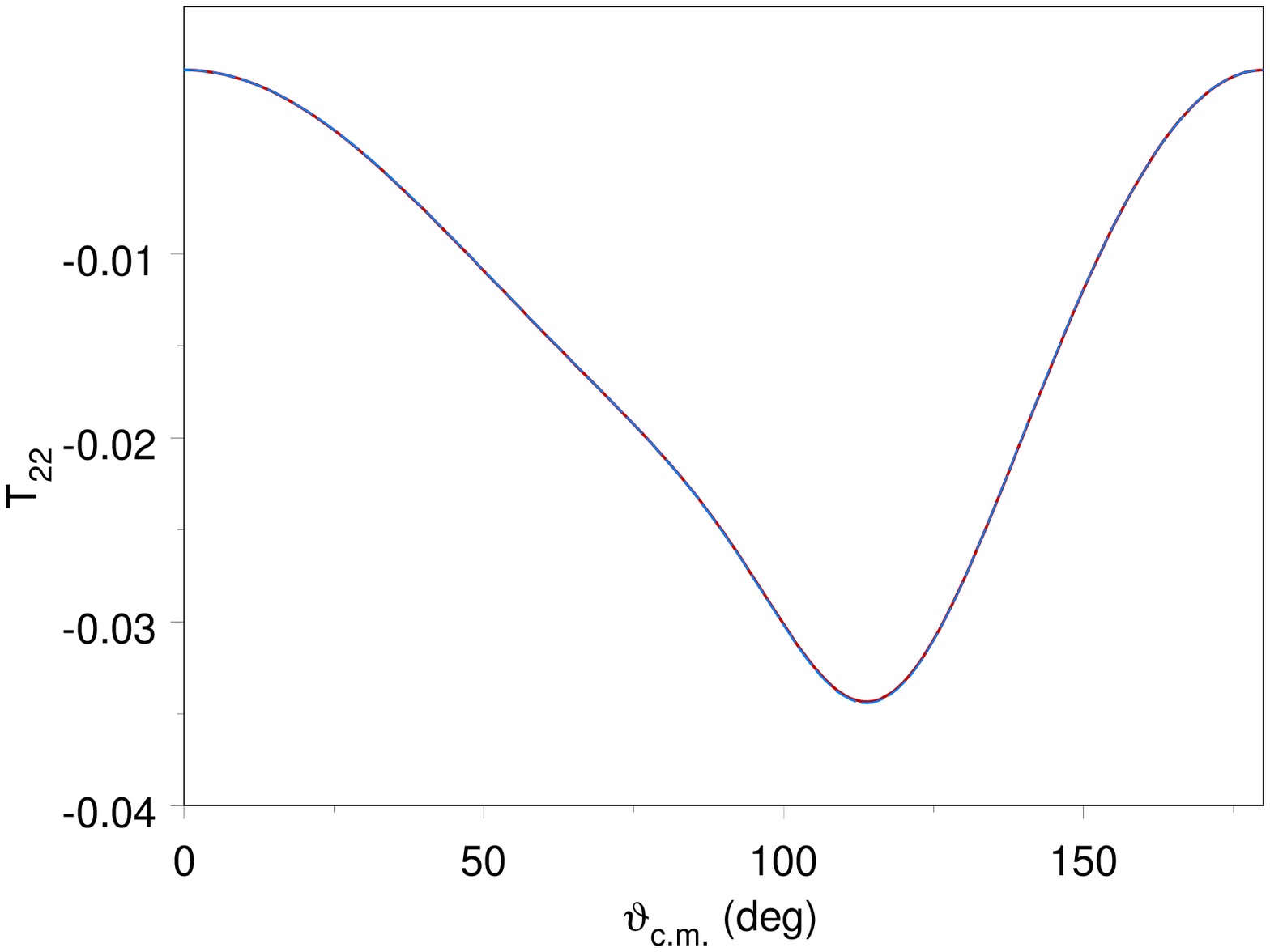}
\end{minipage}
\caption{Deuteron tensor analyzing powers $\rm T_{21}$ and $\rm T_{22}$ for nd elastic scattering at 3~MeV lab
energy as function of the c.m. scattering angle. The notations are the same as in Fig. \ref{fig:1}.}
\label{fig:3}
\end{figure}

Our results for the differential cross section and nucleon analyzing power $\rm A_y$ (Fig. \ref{fig:1}),
deuteron vector $i\rm T_{11}$ and tensor analyzing $\rm T_{20}$ powers (Fig. \ref{fig:2}),
and $\rm T_{21}$ and $\rm T_{22}$ (Fig. \ref{fig:3}) for nd elastic scattering at 3 MeV using
the AV14 NN potential are shown together with the benchmark calculations of Kievsky et al. \cite{KievG}.

     The theoretical predictions are compared with the experimental nd $\rm A_y$ data at 3~MeV \cite{Anin}.
In both calculations all values of the total three-body angular momentum up to M = 15/2 have been used.
In our calculations the total angular momentum of the pair of nucleons $j_{23}$ has been taken up to 3,
while in \cite{KievG} this value was taken up to $j_{23}$ = 4. It should be noted that in the case of
nd scattering increasing $j_{23}$ by unity raises the number of partial waves from 62 up to 98. This difference
presumably explains minor differences between these two calculations around the maximum values
of $\rm A_y$ (Fig. \ref{fig:2}) and of $i\rm T_{11}$ (Fig. \ref{fig:3}), where the predictions of Kievsky et al.
\cite{KievG} are consistently higher by about 2-3 \%. Differences in $\rm T_{20}$, $\rm T_{21}$ and $\rm T_{22}$
are even smaller, about 1 \%.

\begin{figure}[h]
\vfill
\begin{minipage}[h]{100mm}
\includegraphics[width=100mm]{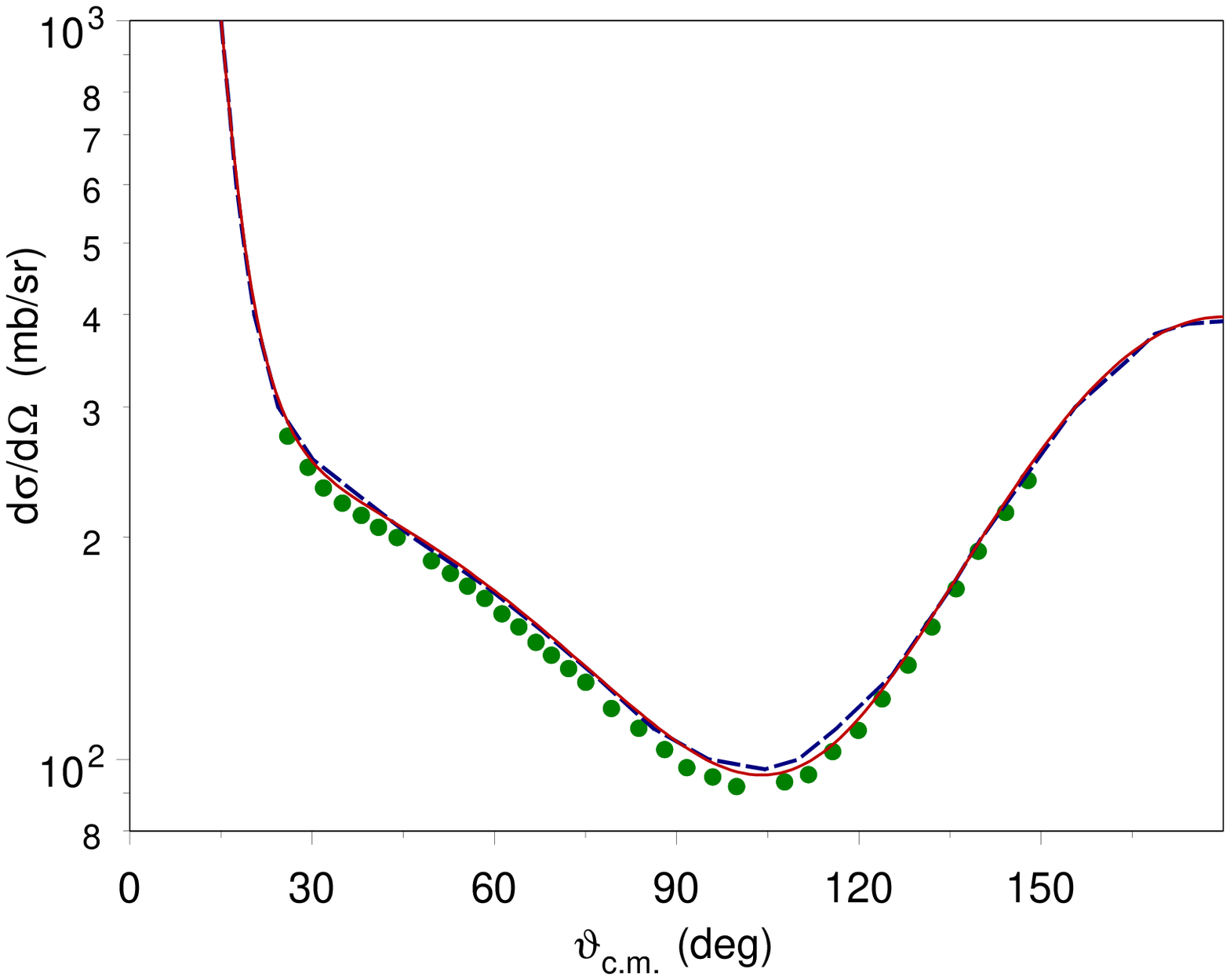}
\end{minipage}
\begin{minipage}[h]{100mm}
\includegraphics[width=100mm]{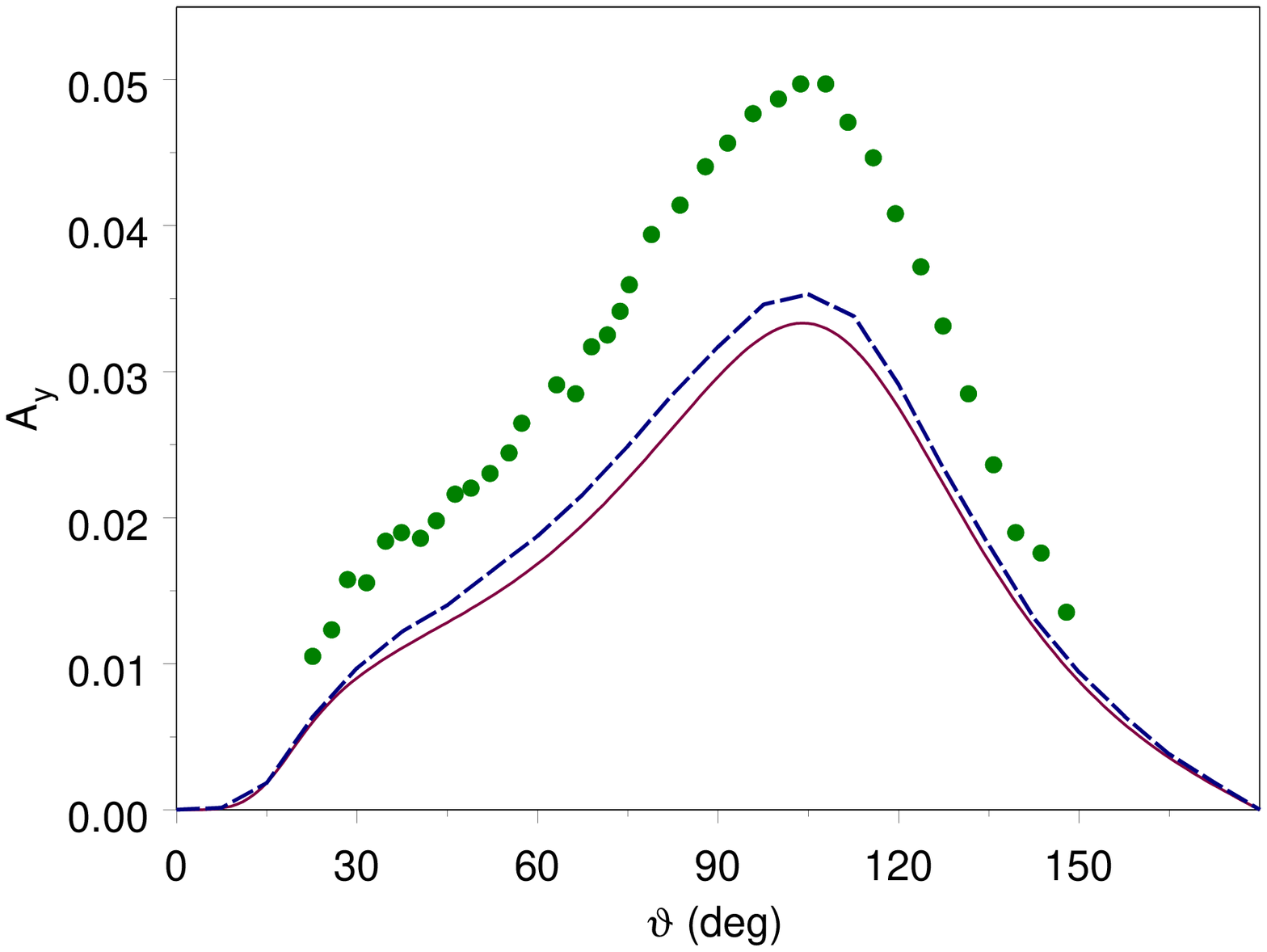}
\end{minipage}
\caption{Differential cross section and proton analyzing power $\rm A_y$ for pd elastic scattering at 3~MeV lab
energy as function of the c.m. scattering angle. The solid lines correspond to our results obtained with AV14 NN
potential. The dashed lines correspond to Deltuva at al. results \cite{DeltK} obtained with AV18 NN potential.
The experimental data are from Ref. \cite{Shim}.}
\label{fig:4}
\end{figure}

\begin{figure}[h]
\vfill
\begin{minipage}[h]{100mm}
\includegraphics[width=100mm]{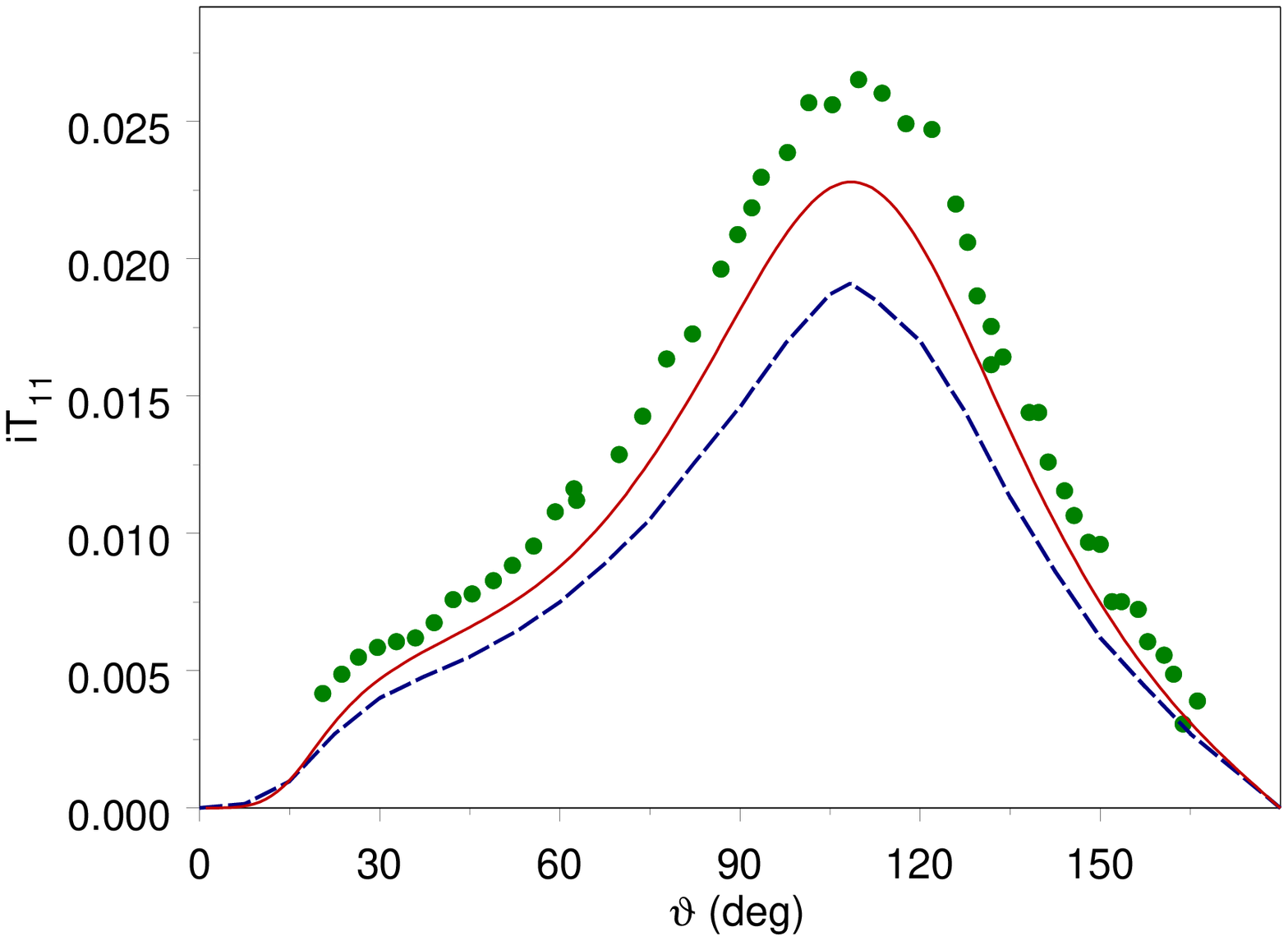}
\end{minipage}
\begin{minipage}[h]{100mm}
\includegraphics[width=100mm]{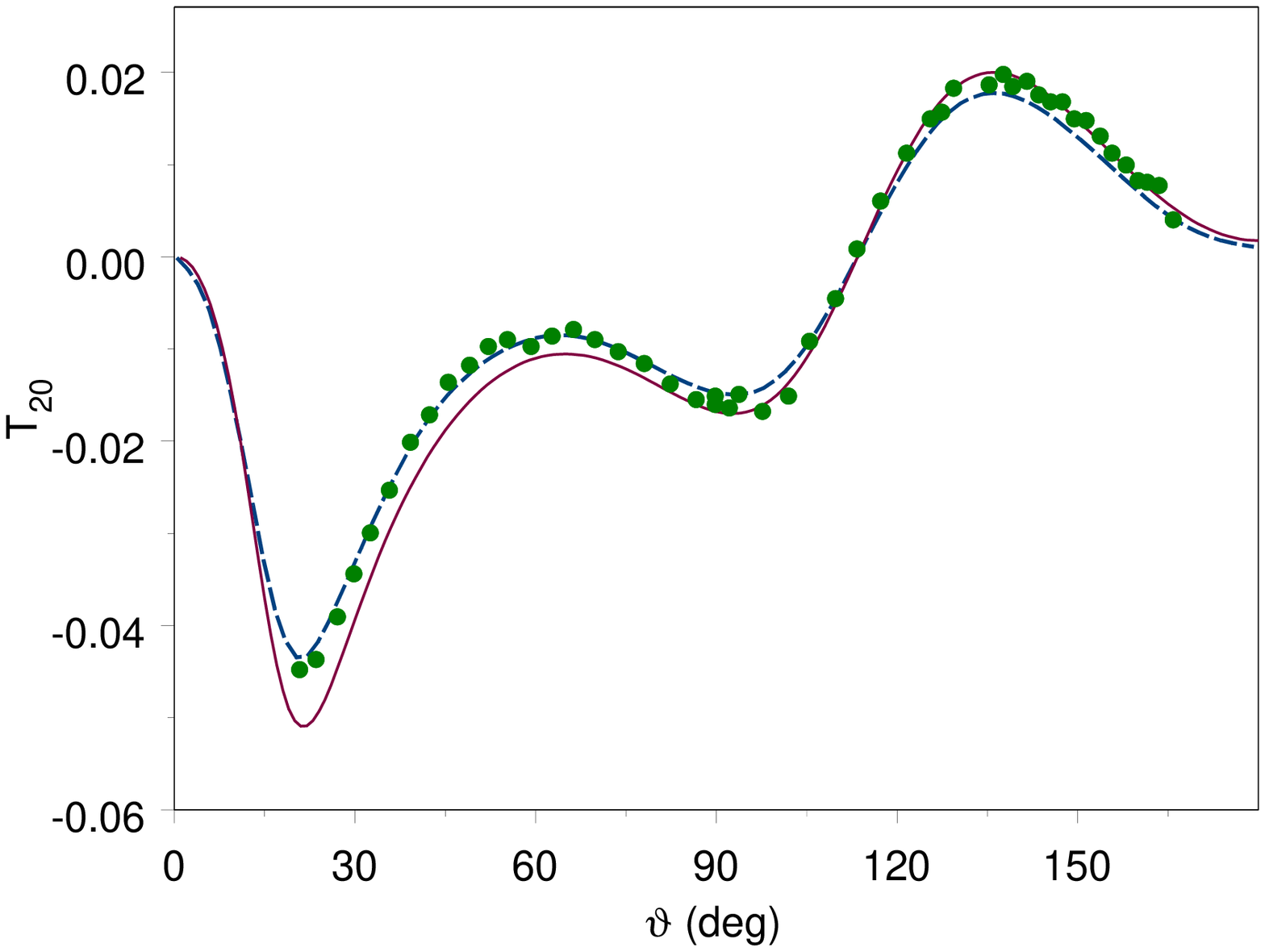}
\end{minipage}
\caption{ Deuteron vector
$i\rm T_{11}$ and tensor $\rm T_{20}$ analyzing power for pd elastic scattering at 3~MeV lab energy as
function of the c.m. scattering angle. The notations are the same as in Fig. \ref{fig:4}.
}
\label{fig:5}
\end{figure}

\begin{figure}[h]
\vfill
\begin{minipage}[h]{100mm}
\includegraphics[width=100mm]{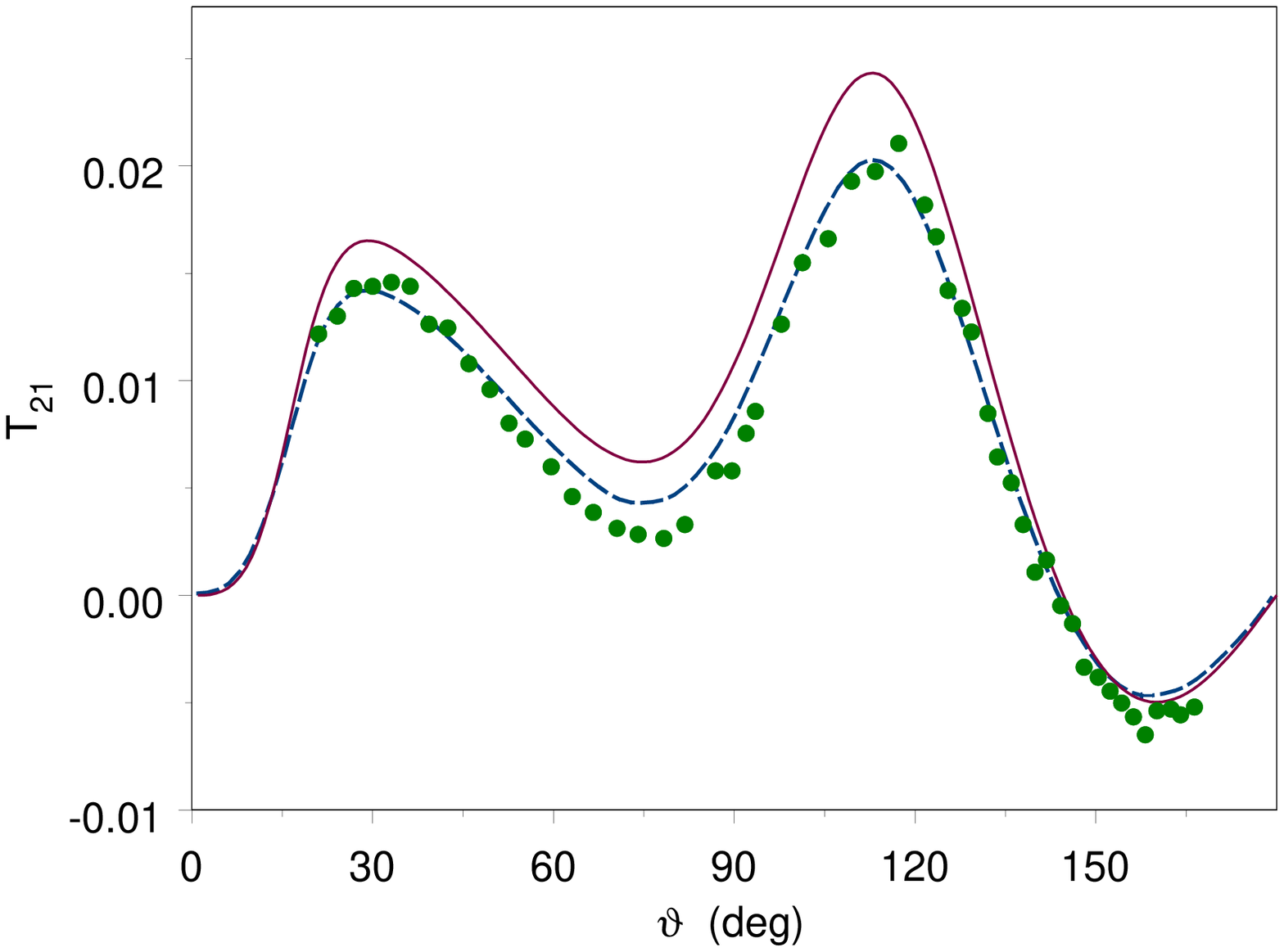}
\end{minipage}
\begin{minipage}[h]{100mm}
\includegraphics[width=100mm]{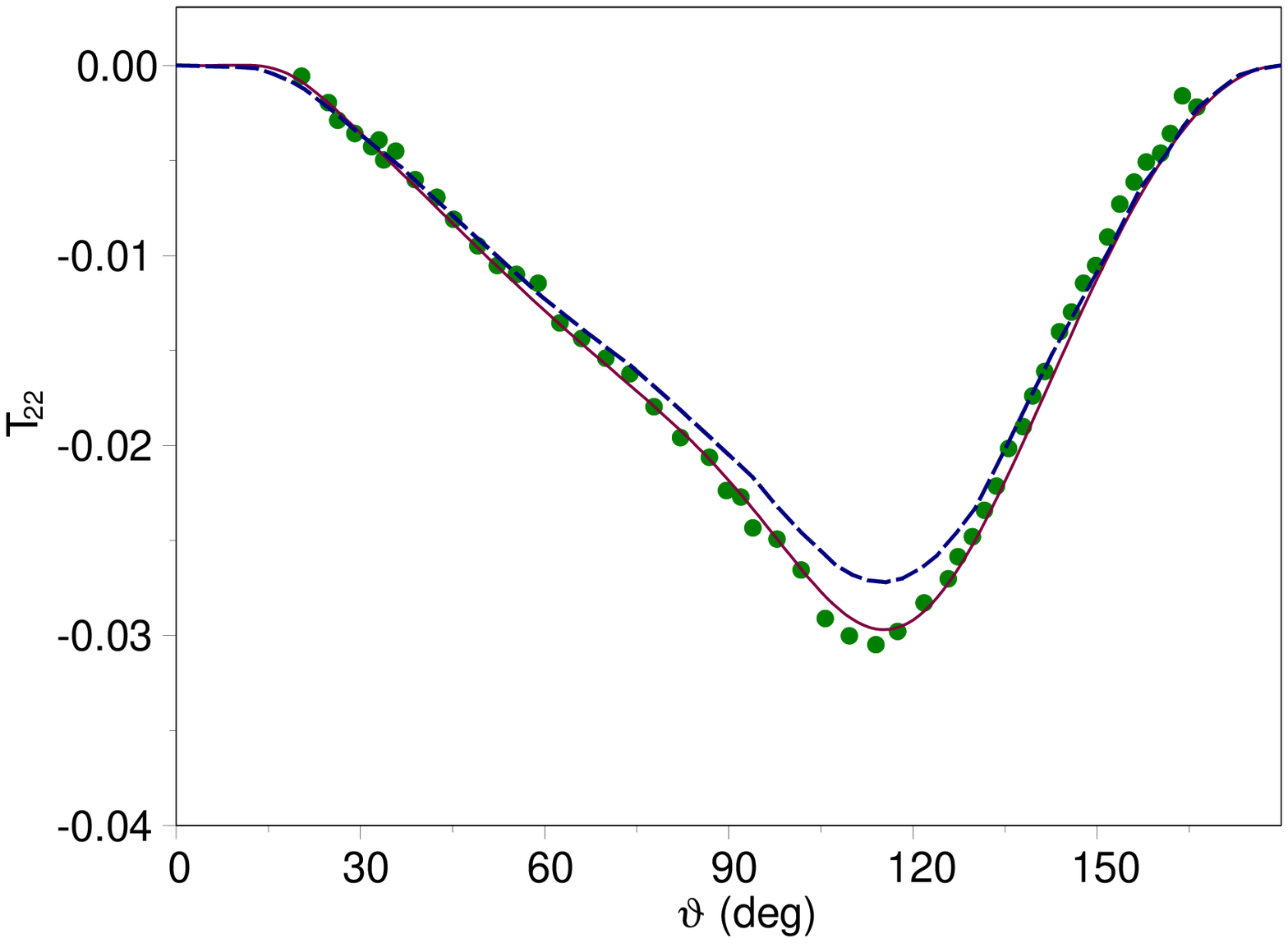}
\end{minipage}
\caption{Deuteron tensor analyzing powers $\rm T_{21}$ and $\rm T_{22}$ for pd elastic scattering at 3~MeV lab
energy as function of the c.m. scattering angle. The notations are the same as in Fig. \ref{fig:4}.}
\label{fig:6}
\end{figure}

     For the pd elastic scattering at 3~MeV, results of our calculations for the differential cross section
and proton analyzing power $\rm A_y$ are shown in Fig. \ref{fig:4} together with those from the benchmark
calculations of Deltuva et al. \cite{DeltK}. Our calculations have been performed using the AV14 NN potential
and involving the correct asymptotic condition to take into account the Coulomb interaction while those of
Ref. \cite{DeltK} used the AV18 NN potential and the screening and normalization procedure for the Coulomb force.
All theoretical calculations are compared with the experimental data of Ref. \cite{Shim}. All values of the
total three-body angular momentum up to M = 15/2 are used in our calculation, while in Ref. \cite{DeltK} value
of M is much larger. We chose values of $j_{23}$ up to 4 (up to 152 partial waves taken into account), whereas
in Ref. \cite{DeltK} these values up to 5 have been used for the strong interaction (207 partial waves were
taken into account). Again this truncation results in a small disagreement between our predictions for
polarization observables and those from Ref. \cite{DeltK}. The results of calculations for the deuteron vector
$i\rm T_{11}$ and tensor $\rm T_{20}$ analyzing powers as well as the experimental data \cite{Shim} are shown
in Fig. \ref{fig:5}. The results of calculations for the deuteron tensor $\rm T_{21}$ and $\rm T_{22}$ analyzing
powers as well as the experimental data \cite{Shim} are shown in Fig. \ref{fig:6}. Predictions of our calculations
and those of Ref. \cite{DeltK} are in reasonable agreement.

     In addition to our new results for nd and pd elastic scattering we would like to present our new results
for pd breakup scattering at $\rm E_{lab}$=14.1 MeV obtained with the Malfliet-Tjon (MT) I-III potential.
In our paper Ref. \cite{Sus} results for inelasticities and phase shifts were obtained in s-wave approximation
for both the strong and the coulomb interactions. This means that only partial waves with $l=0$ were taken into
account for nuclear and electromagnetic forces. It was explicitly pointed out and clearly explained in the paper.
In Table \ref{tab:1} our old and new results together with those of Ref. \cite{DeltK} are given. Our new results
presented in rows $2 - 5$ were obtained for s-wave (MT)I-III potential but the Coulomb interaction was taken
into account for different choices of sets of basis states (number of FNNM equations) in dependence of the maximum
value of the two-body angular momentum $j_{23}$. For the calculation we have neglected by the contribution of
the basis states with the total isospin T=3/2. This negligence makes the relative error less than 0.2\%.
In the Table \ref{tab:1} one can see convergence of our results to those from Ref. \cite{DeltK}. Disagreement
between inelasticity parameters is about 1\% and is about 0.1 degree for phase shifts. As is pointed out in
Ref. \cite{DeltK}, the authors used the perturbation method. Our calculations have been performed by direct
solution of the FNNM equations reduced to a set of linear equations with the resulting matrix having
tri-block-diagonal structure. Small disagreements between results for s-wave pd breakup scattering one can
explain by different numbers of partial waves taken into account (up to 126 in our calculations and up to
398 in calculations of Deltuva et al.). The authors in Ref. \cite{DeltK} have emphasized that for such large
set of basis states direct solution is impossible and one has to apply the perturbation theory.
\section {Discussion}
     Our results for nd elastic scattering at 3~MeV and those from the KVP and momentum-space calculations
are in very good agreement and minor differences can be related to smaller values of $j_{23}$ taken into
account in our calculation. In the energy region from 1.2 to 10~MeV \cite{Neid} theoretical predictions
are 25-30\% lower than the experimental data.

     For pd elastic scattering excellent agreement within 1\% between momentum-space and coordinate-space
calculations based on a variational solution using a correlated hyperspherical expansion predictions at 3, 10
and 65~MeV incident nucleon energies have been demonstrated in Ref. \cite {DeltK}. Predictions of our
calculation and that of Deltuva et al. \cite{DeltK} differ in the use of the NN potential and in values of
the total three-body angular momenta M and of the total angular momenta of the pair of nucleons $j_{23}$
taken into account. Our prediction is about 5\% lower for $\rm A_y$ than that of Ref. \cite{DeltK}, and
surprisingly about 10\% higher for $i\rm T_{11}$. Predictions for tensor analyzing powers agree to better
than 5\%. Comparison with the experimental data of Ref. \cite {Shim} confirms the $\rm A_y$ and $i\rm T_{11}$
puzzles. Both our calculations and those of Ref. \cite{DeltK} are lower than measured values of Ref. \cite{Shim}.
Results using the AV18 NN potential give better agreement with experimental data for $\rm T_{20}$ and $\rm T_{21}$.
However, surprisingly our calculation using the AV14 is in better agreement with the analyzing power $\rm T_{22}$,
possibly indicating differences between AV14 and AV18 potentials.

     To end the discussion, we would like to compare our results for polarization observables with those from
Ref. \cite{Kiev2}. In that paper the authors have performed a detail comparative study of modern three-nucleon
models together in conjunction with the AV18 NN potential to calculate observables for pd elastic scattering
at $\rm E_{lab}$=3~MeV. The authors have shown that only the $\rm N^2\rm{LOL}$ TNF model allows to improve
the description of $\rm A_y$ and $i\rm T_{11}$ noticeably. At the same time the description of $\rm T_{21}$
becomes slightly worse and there is no change in $\rm T_{22}$. In this regard we would like to note that our
predictions obtained with the AV14 NN potential and without three-body forces coincide with the experimental
$\rm T_{22}$ data \cite{Shim} and are in good agreement with the result for $i\rm T_{11}$ from Ref. \cite{Kiev2}
obtained with three-body forces.
\section {Conclusion}
Very good agreement between predictions of our calculations and those of benchmark calculations demonstrates
the soundness of our novel method providing thereby a new approach for calculating three-nucleon scattering
including nucleon-nucleon and electromagnetic interactions. Our approach can and will be used to include three
nucleon forces and to perform additional studies using Kukulin's potential \cite{Kuk} and LS modified
three-nucleon forces of Kievsky \cite{KievLS}, particularly to study the $\rm A_y$ puzzle. It is well-known
that Nd polarization observables are the magnifying glass for studying $^3P_J$ forces and calculations that
rigorously include nuclear and electromagnetic interactions are very valuable.

     Notwithstanding the significance of 3NF, our primary goal is to extend our study using AV14 NN potential
and including the Coulomb potential to energies above the two-body threshold and to focus on breakup data and
on established discrepancies. Our next step is to use the AV18 NN potential. As discussed in this article, we
have already established interesting differences in T20, T21 and T22 most likely due to difference between
AV14 and AV18 NN potentials.
\section{Acknowledgement}
We are very grateful to Prof. H. Witala for courteously giving the computer code to calculate nd observables.
This work was supported by NSF CREST award HRD-0833184 and NASA award NNX09AV07A. The work of I.S. was supported
in part by the Croatian Ministry of Science.

\newpage
\begin{table}
\caption{pd quartet inelasticity and phase shift (in deg.) at $E_{laB}$=14.1 MeV. Nst is the total number of coulomb
partial waves in the case $T$=1/2. For strong interaction only one partial wave ($l$=0) is taken into account.
The results from Ref. \cite{Sus} obtained with a single coulomb partial wave ($l=0$) are given in the first row.
New results obtained for various sets of basis states are given in rows from 2 to 5. The result obtained in
Ref. \cite{DeltK} using perturbation method is given in the last row.}
\label{tab:1}
\begin{ruledtabular}
\begin{tabular}{ccccc}
max$j_{23}$ & Nst & $^4\eta$ & $^4\delta$ \\
\hline
 1        & 1      &0.9202  & 73.64 \\
\hline
 1        & 14     &0.9417  & 73.146 \\
 3        & 46     &0.9682  & 72.744 \\
 4        & 62     &0.9687  & 72.678 \\
 6        & 94     &0.9686  & 72.693 \\
 8        & 126    &0.9686  & 72.696 \\
 \hline
 25       & 398    &0.9795  & 72.604 \\
\end{tabular}
\end{ruledtabular}
\end{table}

\newpage
\newpage
\newpage
\newpage
\newpage
\newpage

\begin{thebibliography}{30}
\bibitem{Gloec}W. Gl\"ockle, H. Witala, D. H\"uber, H. Kamada, J. Golak,
Phys. Rep. {\bf 274}, 107 (1996).
\bibitem{Alt1}E.O. Alt {\it et al}., Phys. Rev. C 17, 1981 (1978); A. Deltuva {\it et al}., Phys. Rev.
C {\bf 72}, 054004 (2005).
\bibitem{Wir1} R. B. Wiringa {\it et al}., Phys. Rev. C {\bf 51}, 38 (1995).
\bibitem{Mach1} R. Machleidt {\it et al}., Phys. Rev. C {\bf 53}, R1483 (1996).
\bibitem{Stoc1} V.G.J. Stocks {\it et al}., Phys. Rev. C {\bf 49}, 2950 (1994).
\bibitem{Wir2} R.B. Wiringa R.A. Smith, and T.L. Ainsworth , Phys. Rev. C {\bf 29}, 1207 (1984).
\bibitem{Coon} S.A. Coon {\it et al}., Nucl. Phys. A {\bf 317}, 242 (1979);
S.A. Coon and H.K. Han, Few-Body Syst. {\bf 30}, 131 (2001).
\bibitem{Pudi}B.S. Pudimer {\it et al}., Phys. Rev. Let. {\bf 51}, 4396 (1995);
S.C. Pieper {\it et al}., Phys. Rev. C {\bf 64}, 014001 (2001).
\bibitem{Mach2} R. Machleidt, Nucl. Phys. A {\bf 790}, 17 (2007), and references therein; U.-G. Meissner,
Nucl. Phys. A {\bf 790}, 129 (2007)
\bibitem{Delt1}A. Deltuva {\it et al}., Phys. Rev. C {\bf 68}, 024005 (2003);
A. Deltuva {\it et al}., Nucl. Phys. A {\bf 790}, 52 (2007).
\bibitem{Navr} P. Navratil, Few-Body Syst. {\bf 41}, 117 (2007).
\bibitem{Ivo} I. Slaus, Nucl. Phys. A {\bf 790}, 199 (2007).
\bibitem{Torn} W. Tornow {\it et al}., Phys. Rev. Lett. {\bf 49}, 312 (1982);
W. Gruebler {\it et al}., Nucl. Phys. A {\bf 398}, 445 (1983); E.M. Neidel {\it et al}., Phys. Lett. B {\bf 552}, 29 (2003), and references therein.
\bibitem{Wiel} B.J. Wielinga {\it et al}., Nucl. Phys. A {\bf 261}, 13 (1976);
H.R. Setze {\it et al}., Phys. Lett. B {\bf 388}, 229 (1996), and references therein; Z. Zhou {\it et al}.,
Nucl. Phys. A {\bf 684}, 545 (2001); J. Ley {\it at al}., Phys. Rev. C {\bf 73}, 064001 (2006).
\bibitem{How} C.R. Howel {\it et al}., submitted to Phys. Rev. C.
\bibitem{Kiev1} A. Kievsky {\it et al}., Nucl. Phys. A 577, 511 (1994);
A. Kievsky {\it et al}., Phys. Rev. C {\bf 58}, 3085 (1998); A. Kievsky {\it et al}., Phys. Rev. C {\bf 64}, 024002 (2001);
A. Kievsky {\it et al}., Phys. Rev. C {\bf 69}, 014002 (2004).
\bibitem{Alt2} E.O. Alt {\it et al}., Nucl. Phys. B {\bf 2}, 167 (1967).
\bibitem{Kiev2} A. Kievsky, nucl-th ArXiv 1002.1254, Feb 5, 2010 and arXiv 1002.1601 Feb 8, 2010 and references therein.
\bibitem{MGL} S.P. Merkuriev, C. Gignoux and A. Laverne, Ann. Phys. {\bf 99}, 30 (1976).
\bibitem{KKM} A.A. Kvitsinsky, Yu.A. Kuperin, S.P. Merkuriev, A.K. Motovilov
and  S.L. Yakovlev, Fiz. Elem. Chastis At. Yadra {\bf 17}, 267 (1986).
\bibitem{MerkAs} S.P. Merkuriev, Ann. Phys. (N.Y.) {\bf 130}, 3975 (1980),\\
S.P. Merkuriev, Acta Physica (Austriaca), {\bf Suppl. XXIII}, 65 (1981).
\bibitem{Sus}  V.M. Suslov and B. Vlahovic, Phys. Rev. C{\bf 69}, 044003 (2004).
\bibitem{KviH} A.A. Kvitsinsky and C.-Y. Hu, Few-Body Syst. {\bf 12}, 7 (1992).
\bibitem{KievG} A. Kievsky, M. Viviani, S. Rosati, D. H\"uber, W. Gl\"ockle, H. Kamada, H. Witala, and J. Golak,
Phys. Rev. C {\bf 58}, 3085 (1998).
\bibitem{Anin} J.E. McAninch, W. Haeberli, H. Witala, W. Gl\"ockle and J. Golak, Phys. Lett. B {\bf 307}, 13 (1993).
\bibitem{DeltK} A. Deltuva, A.C. Fonseca, A. Kievsky, S. Rosati, P.U. Sauer, and M. Viviani, Phys. Rev.
C {\bf 71}, 064003 (2005).
\bibitem{Shim} S. Shimizu, K.Sagara, H. Nakamura, K. Maeda, N. Nishimori, S. Ueno, T. Nakashima, and S. Morinobu,
Phys. Rev. C {\bf 52}, 1193 (1995).
\bibitem{Neid} E.M. Neidel {\it et al}., Phys. Lett. B {\bf 552}, 29 (2003).
\bibitem{Kuk} V.I. Kukulin, 18th International IUPAP Conference
on Few-Body Problems in Physics, FB18 - Book of Abstracts, p.215 (2006).
\bibitem{KievLS} A. Kievsky, Phys. Rev. C {\bf 60}, 031001 (1999).
\end{thebibliography}
\end{document}